# Parameterization of Cherenkov Light Lateral Distribution Function as a Function of the Zenith Angle around the Knee Region

Marwah M. Abdulsttar[a], A. A. Al-Rubaiee[b, *], Abdul Halim KH. Ali[c]

College of Science, Dept. of Physics, Al-Mustansiriyah University,

10052 Baghdad, Iraq

[a]gonti77@mail.ru, [b]dr.ahmedrubaiee@gmail.com, [c]halimkh@yahoo.com

**Keywords:** Extensive Air Showers, Cherenkov radiation, lateral distribution function, zenith angle.

**Abstract.** Cherenkov light lateral distribution function (CLLDF) simulation was fulfilled using CORSIKA code for configurations of Tunka EAS array of different zenith angles. The parameterization of the CLLDF was carried out as a function of the distance from the shower core in extensive air showers (EAS) and zenith angle on the basis of the CORSIKA simulation of primary proton around the knee region with the energy $3.10^{15}$ eV at different zenith angles. The parameterized CLLDF is verified in comparison with the simulation that performed using CORSIKA code for two zenith angles.

1. **Introduction**

The effect of the Cherenkov light can be seen when charged particles breaks through dielectric medium, like air, exceeds the light's phase in the medium ($v > c/n$), when $c$ is the speed of light; $v$ is the speed of the charged particles and $n$ is the refractive index. The Cherenkov light emission can be described by the superposition of spherical waves using Huygen's principle [1]. In Fig. 1 is demonstrated the radiation that observed only in a distinguished cone with a fixed angle $\theta$ between the emission direction and the particle velocity where the Cherenkov angle can be given as [1]:

$$\cos\theta = \frac{(c/n)t}{(\beta c)t} = \frac{1}{\beta n} \qquad (1)$$

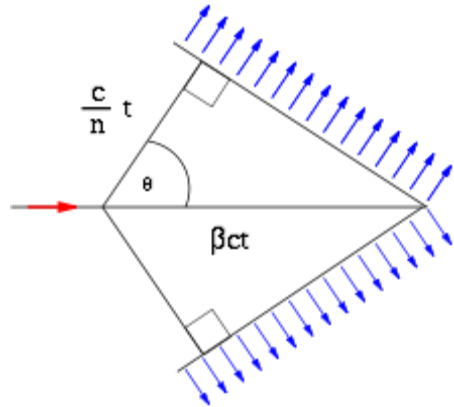

**Fig. 1** Huygen's construction of the Cherenkov radiation [1].

E. Korosteleva et al. proposed CLLDF contains a single shape parameter that depended only upon the relative position of the shower maximum for any primary nucleus, primary energy and arrival direction from the data of QUEST experiment for the EAS-TOP array and the CORSIKA simulation code [2]. S. Cht. Mavrodiev et al. proposed a new method for estimation of the mass composition and the energy spectrum of primary cosmic radiation based only on atmospheric Cherenkov light flux analysis [3].

In the present work, the simulation of the CLLDF for the primary proton using CORSIKA code for the configurations of the Tunka-133 EAS array around the knee region with the energy $3.10^{15}$ eV at different zenith angles was fulfilled. Through this simulation, a fitting function of the CLLDF with its parameters as a function of the distance from the shower axis $R$ and zenith angle $\theta$ it was carried out. The obtained parameterized function has been verified in comparison with the CORSIKA code for two fixed zenith angles $22^o$ and $42^o$ at the primary energy 3 PeV.

## 2. Tunka-133 Cherenkov array

Tunka-133 array consists of 133 wide-angle optical detectors on the basis of PMT EMI 9350 with a hemispherical photocathode of 20cm diameter. Tunka-133 is located in the Tunka valley (in Russia) with about 1 km$^2$ geometric area designed to record EAS from CRs or high energy gamma rays and detect the Cherenkov radiation of air showers during dark and clear nights [4, 5]. The optical detectors in the array separated to 19 compact sub arrays called clusters, each cluster composed of 7 optical detectors in each one-six hexagonally arranged detectors and one in the center [6]. The distance between the detectors is 85 m and the configuration of the array is shown in Fig. 2 [7].

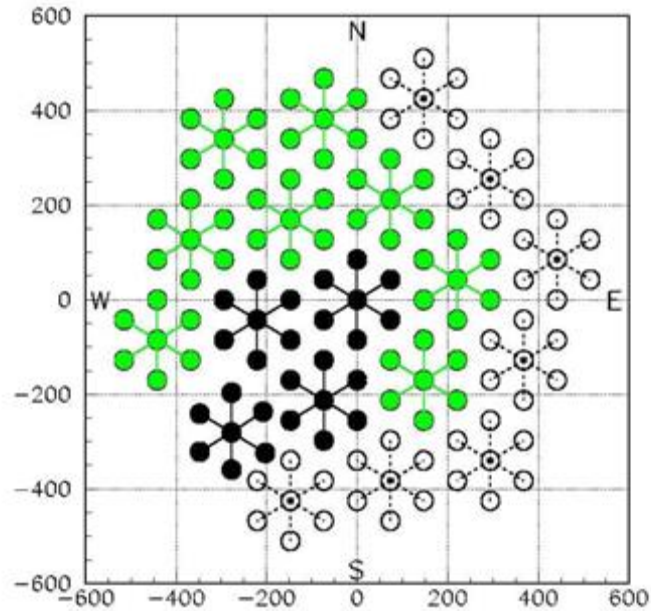

**Fig. 2** Tunka-133 detector configurations [7].

## 3. Results and Discussion

### 3.1 CLLDF simulation using CORSIKA package

Cosmic Ray Simulations for KAscade (CORSIKA) code is a detailed Monte Carlo program for studying the evolution and properties of EAS in the atmosphere [8]. In the present work, the simulation of CLLDF is performed using CORSIKA code by using two hadronic models: QGSJET (Quark Gluon String model with JETs) code [9] that was used to model interactions of hadrons with energies greater than 80 GeV and GHEISHA (Gamma Hadron Electron Interaction Shower) code [10] which is used for energies lower than 80 GeV. The simulation of CORSIKA program was performed for the configuration of the Tunka-133 for primary proton with the energy 3.10¹⁵ eV at different zenith angles [11]. In Fig. 3 it was demonstrated the simulation of CLLDF for primary proton and different zenith angles (0°- 45°) at fixed primary energy (3.10¹⁵ eV). Through this figure one can see that the CLLDF is inversely proportional with zenith angle and $R$ i.e. the Cherenkov radiation decreases with increasing the zenith angle and $R$.

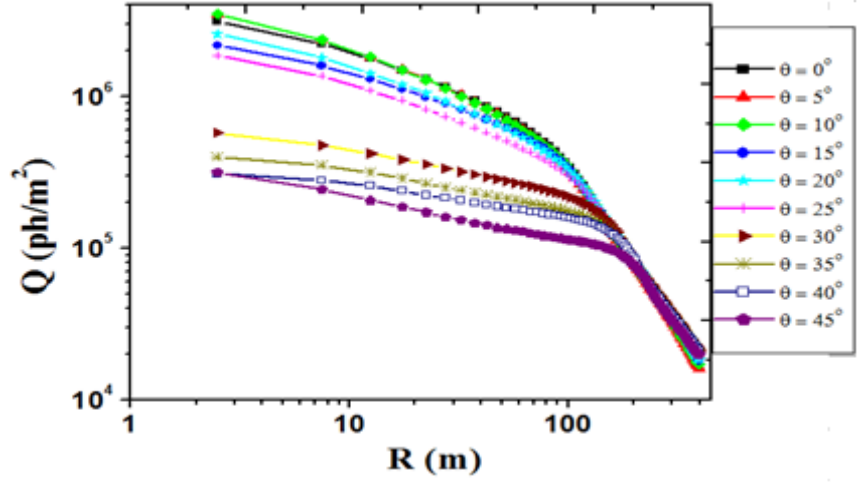

**Fig. 3** The CLLDF data results using CORSIKA simulations of primary proton with the energy 3.10¹⁵ eV at different zenith angles.

### 3.2 Parameterization of the CLLDF

The number of Cherenkov photons per the interval of wavelength ($\lambda_1$, $\lambda_2$) can be obtained from the Tamm and Cherenkov expression [12]:

$$\frac{dN_\gamma}{dx} = 2\pi\alpha \sin^2\theta_r \int_{\lambda_1}^{\lambda_2} \frac{d\lambda}{\lambda^2} =$$
$$= 2\pi\alpha \left(\frac{1}{\lambda_1} - \frac{1}{\lambda_2}\right) \zeta_o \left(1 - \frac{E_{th}^2(h)}{E^2}\right) \exp(-h/h_o), \qquad (2)$$

where $\alpha = 1/137$ is the fine structure constant; $\zeta_0 \approx 3.10^{-4}$, $h_0 = 7.5\ Km$; $E_{th} = mc^2\gamma_{th}$, which is the threshold energy of electron at the atmosphere at the height $h$; $\gamma_{th}$ is the Lorenz factor at the threshold energy which can be defined as:

$$\gamma_{th} = \frac{1}{\sqrt{1-\beta^2}} = \frac{1}{\sqrt{1-(1/n(h))^2}} \qquad (3)$$

At sea level $n = 1 + \zeta_o$, and the Cherenkov light can be emitted by the electrons which exceed $E_{th}$ i.e. when $\gamma > \gamma_{th} = \frac{E_{th}}{mc^2} \approx 40.8$. By neglecting the absorption of Cherenkov light in the atmosphere, the total number of Cherenkov photons $N_\gamma$ that radiated by electrons will be written as:

$$N_\gamma = 45.10^{10} \frac{E_o}{10^{15}ev} \qquad (4)$$

Estimation of the core position and age parameter are also made by using the total number of Cherenkov photons in EAS, which is directly proportional to primary energy ($E_0$) [13]:

$$N_\gamma = 3.7 \cdot 10^3 \frac{E_o}{\beta_t}, \qquad (5)$$

where $\beta_t$ is the critical energy at which it equals ionization losses at the t-unit: $\beta_t = \beta_{ion} t_o$. For electron, $\beta_{ion} = 2.2$ Mev. $(g \cdot cm^{-2})^{-1}$, $t_o = 37$ g. $cm^{-2}$ and $\beta_t = 81.4$ MeV [14].

Thus, the number of Cherenkov photons in the shower is directly proportional to the primary particle energy. Experimentally, this value it is not easy to measure; therefore, the density of the Cherenkov light, namely, the number of photons per unit area of the detector not model related [15] to the primary energy of particles:

$$Q_{(E,R)} = \frac{\Delta N_{\gamma\,(E,R)}}{\Delta S}, \qquad (6)$$

is used for experimental data processing. As demonstrated by direct measurements of Cherenkov radiation [14].

The parameterization of the simulated CLLDF was performed with a polynomial expression as a function of the distance $R$ from the shower axis which depended on five parameters a, b, c, d and e:

$$\log[Q(R)] = a + b\log(R) + c\log^2(R) + d\log^3(R) + e\log^4(R) \qquad (7)$$

Fig. 4 shows the fitting curves of the CLLDF as a function of the distance $R$ from the shower axis for primary proton with the energy $3 \cdot 10^{15}$ eV for Tunka-133 EAS array for different zenith angles.

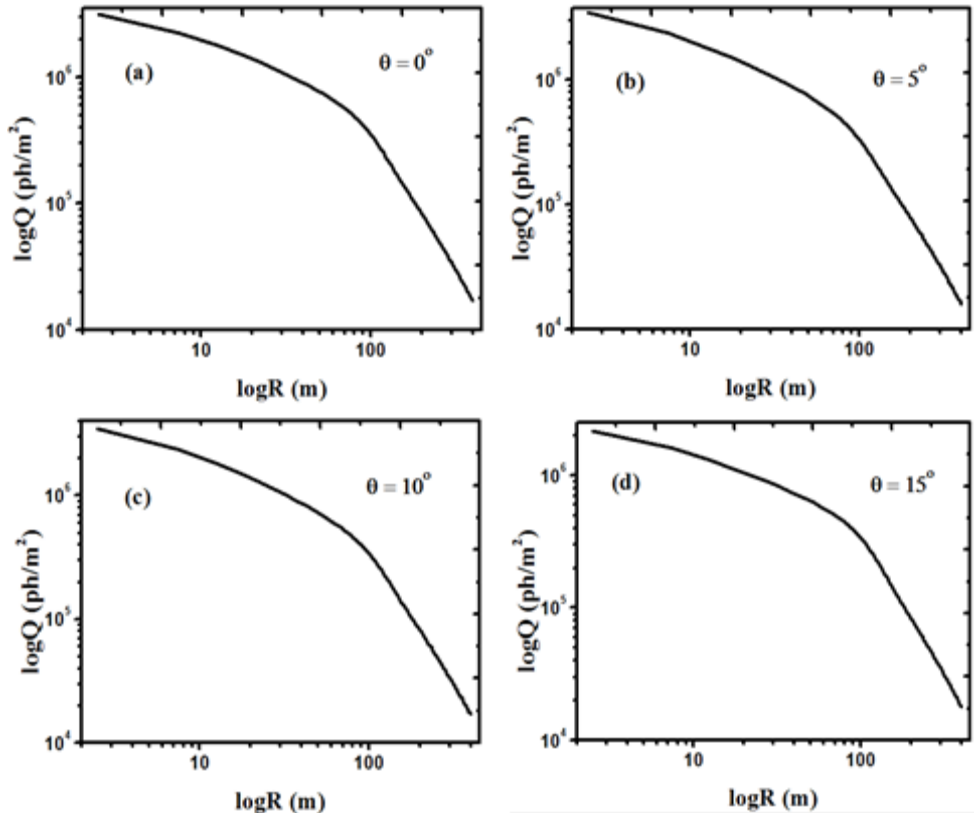

(Continued with Fig. 4)

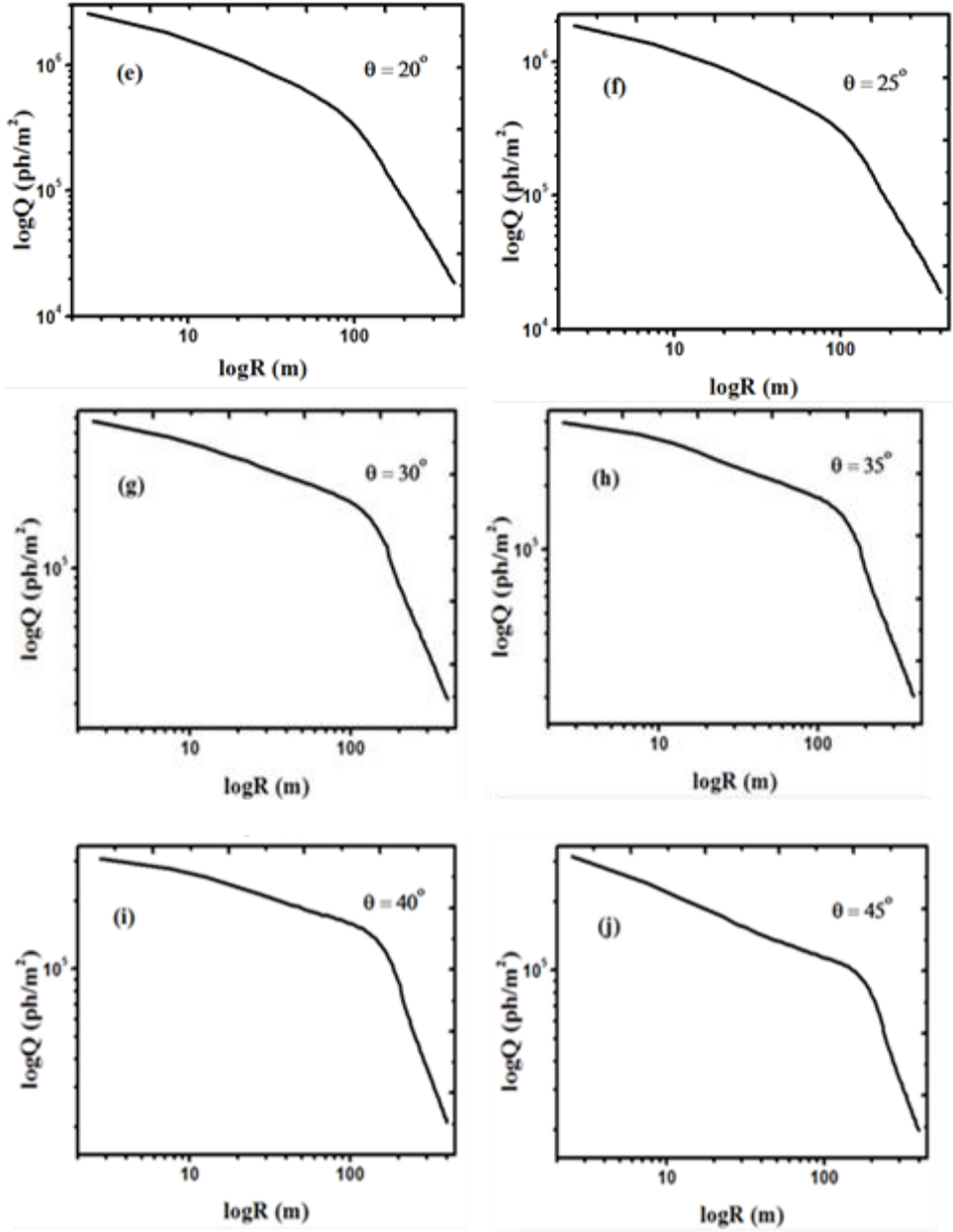

**Fig. 4** Fitting curves of the CLLDF as a function of the distance $R$ from the shower axis of primary proton with the energy $3.10^{15}$ eV at different zenith angles.

The parameterization of the five parameters of Eq. 7 was performed as a function of the zenith angle with the following expressions:

$$a = 6.548 + 0.1515\theta - 0.0083\theta^2 + 9.11 \times 10^{-5}\theta^3 \tag{8}$$

$$b = -1.29 - 0.29\theta + 0.0138\theta^2 - 1.2133 \times 10^{-4}\theta^3 \tag{9}$$

$$c = 1.95 + 0.224\theta - 0.01175\theta^2 + 8.313 \times 10^{-5}\theta^3 \tag{10}$$

$$d = -1.5744 - 0.00545\theta + 0.00182\theta^2 + 5.483 \times 10^{-6}\theta^3 \tag{11}$$

$$e = 0.32 - 0.01468\theta + 2.9511\times10^{-4}\theta^2 - 7.8197\times10^{-6}\theta^3 \qquad (12)$$

The polynomial fitting curves of the parameters a, b, c, d and e of the CLLDF function (Eq. 7) as a function of zenith angles are shown in Fig. 5.

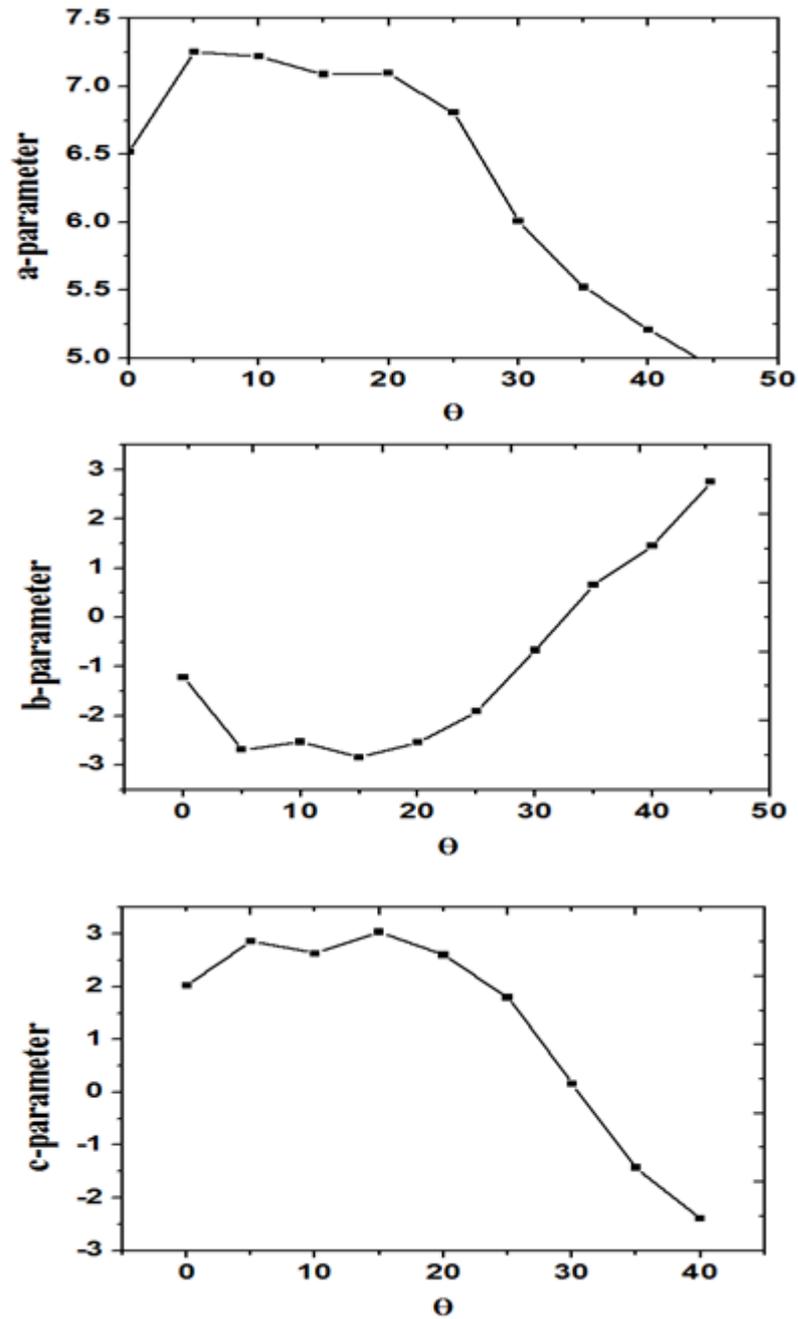

(Continued with Fig. 5)

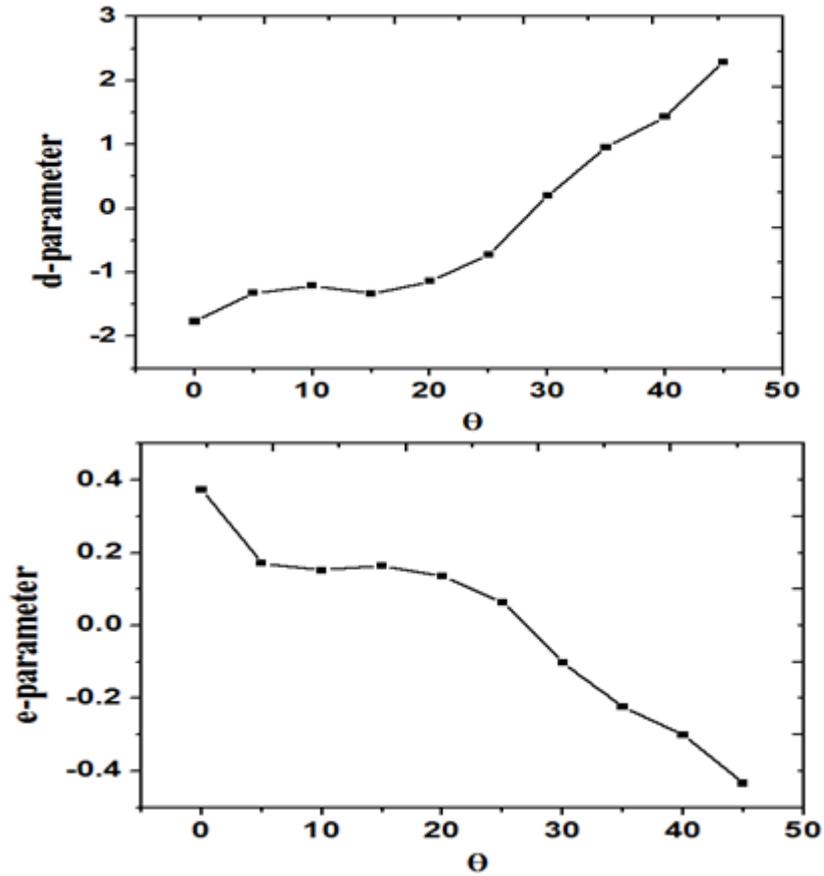

**Fig. 5** Dependence of the parameters a, b, c, d, and e of Eq. 7 on the zenith angle.

Equation 7 with its parameters (Eqs. 8-12) is a function of the distance *R* from the shower core and zenith angle, which are verified in comparison with the CLLDF which was simulated using CORSIKA code for two zenith angles 22° and 42° for the fixed energy 3 PeV as shown in Fig. 6.

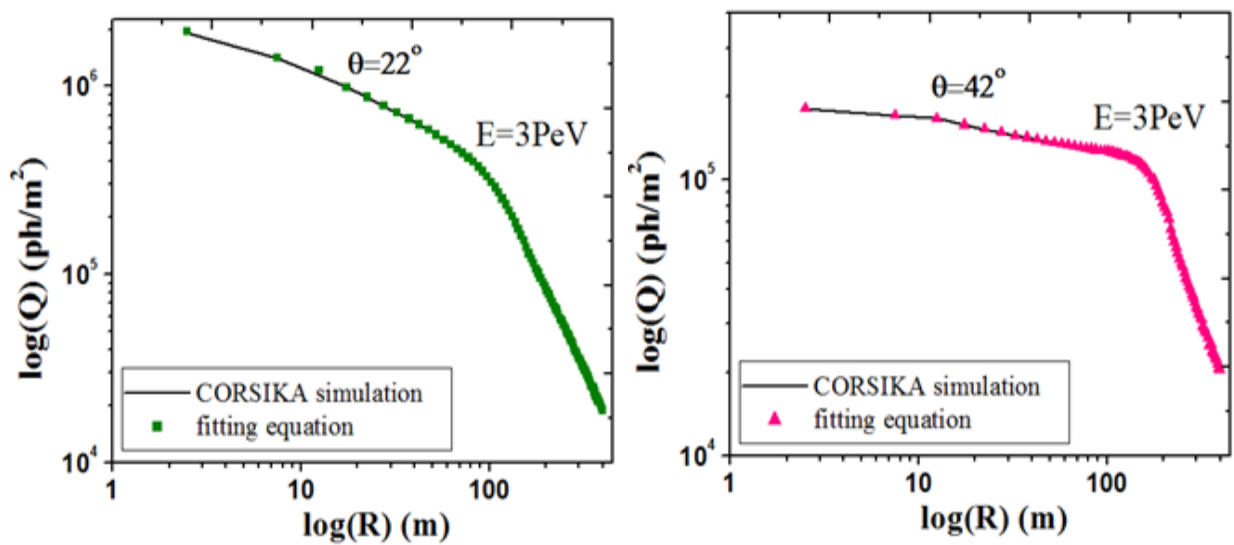

**Fig. 6** Verification of the parameterized CLLDF parameters (Eq. 7) with CORSIKA code simulation for primary proton for two different zenith angles: (a) θ = 22° and (b) θ = 42°.

Through the equations 7-12, one can see that the parameterization of the CLLDF was performed as a function of the distance from the shower core in EAS and zenith angle on the basis of the CORSIKA simulation of primary proton around the knee region with the energy $3.10^{15}$ eV at different zenith angles. The parametrized CLLDF was verified in comparison with that simulated CLLDF for two fixed zenith angles $22^o$ and $42^o$. The fundamental advantage of the given approach consists of the ability to make a library of CLLDF samples that could be used to analyzing of real events which can be detected with the EAS arrays and reconstruction of the primary CRs energy spectrum and mass composition.

### 4. Conclusion

A new fitting function with its parameters for the Cherenkov light lateral distribution function as a function of the distance from the shower axis $R$ and zenith angle $\theta$ was obtained. The parameterized function can be used to estimate the Cherenkov light lateral distribution function for primary proton around the knee region of the cosmic ray energy spectrum as a function of the distance from the shower core and the zenith angle. This parameterized function is already verified in comparison with the lateral functions of Cherenkov radiation that were simulated using CORSIKA program for two fixed zenith angles $22^o$ and $42^o$ around the knee region of cosmic ray energy spectrum.